\documentstyle[twoside,fleqn]{article}
\newcommand{\bls}[1]{\renewcommand{\baselinestretch}{#1}}

\def\twocol{\twocolumn \mathindent=1em}
\def\noi{\noindent}

\newcommand{\psection}[1]{{\raggedright\section{\protect\large\bf #1}}}

\newcommand{\psubsection}[1]%
   {{\raggedright\subsection{\protect\normalsize\bf #1}}}

\renewcommand{\thesubsubsection}%
   {\arabic{section}.\arabic{subsection}.\arabic{subsubsection}.}
\newcommand{\heads}[2]{\markboth{\protect\small\it #1}{\protect\small\it #2}}
\newcommand{\Acknow}[1]{\subsection*{Acknowledgement} #1}

\newcommand{\Title}[1]{\noindent {\Large #1} \\}
\newcommand{\Author}[2]{\noindent{\large\bf #1}\\[2ex]\noindent{\it #2}\\}

\newcommand{\foom}[1]{\protect\footnotemark[#1]}

\newcommand{\email}[2]{\footnotetext[#1]{e-mail: #2}}

\newcommand{\WFighere}[3]
       {\framebox[\textwidth]{\rule{0cm}{#2}} \vspace{-3pt} \\
        Figure #1: {\small #3 } \smallskip\hrule\bigskip
        \addtocounter{figure}{1}}

\def\nq{\hspace{-1em}}
\def\nqq{\hspace{-2em}}

\def\beq{\begin{equation}}
\def\eeq{\end{equation}}
\def\bear{\begin{eqnarray}}

\def\lal{&&\nqq {}}               
\def\bearr{\begin{eqnarray} \lal}
\def\ear{\end{eqnarray}}
\def\earn{\nonumber \end{eqnarray}}

\def\tst{\textstyle}

\def\nn{\nonumber\\ {}}

\def\nnn{\nonumber\\ \lal }

\def\yy{\\[5pt]}

\def\e{{\,\rm e}}

\def\half{{\tst\frac{1}{2}}}

\newcommand{\aver}[1]{\langle \, #1 \, \rangle \mathstrut}

\catcode`\@=11 \@addtoreset{equation}{section}\catcode`\@=12
\newcommand{\be}{\begin{equation}}
\newcommand{\ee}{\end{equation}}
\newcommand{\ba}{\begin{eqnarray}}
\newcommand{\ea}{\end{eqnarray}}
\newcommand{\p}{\partial}
\newcommand{\btd}{\bigtriangledown}
\newcommand{\btu}{\bigtriangleup}
\newcommand{\sq}[1]{\sqrt{|#1|}}
\newcommand{\avers}[1]{{\aver{#1}}_*}

\heads{V.D. Ivashchuk and V.N. Melnikov}
{Intersecting $p$-brane Solutions in Multidimensional Gravity and
M-theory}

\bls{1.01}
\begin{document}
\setcounter{page}{0}
\thispagestyle{empty}
\large
\begin{center}
               RUSSIAN GRAVITATIONAL SOCIETY\\
               INSTITUTE OF METROLOGICAL SERVICE \\
               CENTER OF GRAVITATION AND FUNDAMENTAL METROLOGY\\
\end{center}
\vskip 4ex
\begin{flushright}
                                         RGS-VNIIMS-004-96\\
                                         hep-th/9612089
\end{flushright}
\vskip 25mm
\begin{center}

\Title{\Large\bf INTERSECTING P-BRANE SOLUTIONS \yy
                 IN MULTIDIMENSIONAL GRAVITY AND M-THEORY}

\bigskip

\Author{V.D. Ivashchuk\foom 1 and V.N. Melnikov\foom 2}
{Center for Gravitation and Fundamental Metrology, VNIIMS,\\
     3--1 M. Ulyanovoy Str., Moscow 117313, Russia}

\end{center}

\bigskip
\noi     {\bf Abstract}
\medskip

\noi
Multidimensional gravitational model on the manifold
$M = M_0 \times \prod_{i=1}^{n} M_i$, where  $M_i$
are Einstein spaces ($i \geq 1$), is considered. The action contains
$m = 2^n -1$ dilatonic scalar fields $\varphi^I$ and $m$  (antisymmetric)
 forms $A^I$.  When all fields and scale factors of the metric
depend (essentially) on the point of $M_0$ and any $A^I$ is "proportional"
to the volume form of submanifold
$M_{i_1}  \times  \ldots \times M_{i_k}$, $1 \leq i_1 < \ldots < i_k \leq
n$, the $\sigma$-model representation is obtained.  A family of
"Majumdar-Papapetrou type" solutions are obtained, when all $M_{\nu}$ are
Ricci-flat. A special class of solutions (related to the solution of some
Diophantus equation on dimensions of $M_{\nu}$)  is singled out. Some
examples of intersecting $p$-branes (e.g. solution with seven Euclidean
$2$-branes for $D = 11$ supergravity) are considered.

\vfill

{\small \noi Submitted to {\it Gravitation and Cosmology}}
\vspace {1cm}

\centerline{Moscow 1996}]

\email 1 {ivas@cvsi.rc.ac.ru}
\email 2 {mel@cvsi.rc.ac.ru}

\twocol
\normalsize

\psection{Introduction}

Recently a growth of interest to $D=11$ sypergravity  \cite{CJS,SS}
arouse. This growth was stimulated mainly  by a suggestion that
there exists a new  "fundamental" theory, so-called "M-theory"
\cite{HTW}-\cite{D}, which generalizes all known string theories
\cite{GSW} and whose low energy limit is
$D=11$ supergravity. $M$-theory describes (amongst other things) two
supersymmetric extended objects: supermembrane (with two spatial dimensions)
and superfivebrane (with five spatial dimensions).
There exists also a conjecture about existing of other fundamental
12-dimensional theory, so-called $F$-theory \cite{Vafa}.

Exact solutions, especially so-called $p$-brane solutions
\cite{Dab}-\cite{AVV} play a  rather important role  in investigation of
superstrings in various dimensions and its relation to $M$-theory.

A simplest $p$-brane solution has the following form
\cite{Dab}, \cite{LPSS}-\cite{GHT}
\beq
\label{1.1}
g= H^{\alpha} [ \delta_{ab} dx^{a} \otimes dx^{b} +
   H^{\beta} \eta_{m n} dy^{m} \otimes dy^{n}] ,
\eeq
where $m, n = 0, \ldots, p$; $\alpha$, $\beta$
are some parameters and $H = H(x)$ is harmonic function.
Here $p=1$ corresponds to string, $p=2$ to membrane etc. For $p=0$ in
(1.1) we usually get an extremal black hole or multicenter
generalization a la Majumdar-Papapetrou \cite{MP}. The solution (\ref{1.1})
occurs in the model with $(1 + p)$-form $A$ "proportional" to the volume
form on ${\bf R}^{1+p}$.

In this paper we consider the generalization of the solutions (\ref{1.1})
to multimembrane case for arbitrary dimension, when the action
of the model  (see (\ref{2.1}) below)
contains metric,  set of dilatonic scalar
fields $\varphi^I$ and (antisymmetric) forms $A^I$, $I \in \Omega$,
interacting by exponential dilatonic coupling.
Actions of such type effectively occur in various supergravity models
\cite{CJS,SS,LPS1}, when certain classes of pure bosonic solutions are
considered. Here we consider the "multidimensional space-time"
in the product form $M = M_0 \times \prod_{i=1}^{n} M_i$, where  $M_0$
is "our space" and $M_{i}$ ($i \geq 1$) are
internal (originally Einstein) spaces.
All manifolds $M_{\nu}$, $\nu \geq 0$ are oriented and connected.
We say that the submanifold
$M_I = M_{i_1}  \times  \ldots \times M_{i_k}$,
$1 \leq i_1 < \ldots < i_k \leq n$, forms a $p$-brane if there exists
a solution with a non-zero form $A^I$ "proportional" to
the volume form of $M_I$ (see (2.17)). We  consider the simplest
case when scale factors of the metric and all fields
depend (essentially) on the point of $M_0$
, i.e. the isotropic case is considered. The solutions
of  such type  (mainly with flat internal spaces $M_i$)
governed by several harmonics functions
are intensively studied in literature (see, for example,
\cite{PT}-\cite{LPS2}). But, to our knowledge, authors
do not use  regular (general) schemes for obtaining
such solutions. For example, in  \cite{Ts1} a lot of interesting
overlapping supersymmetric  $p$-brane solutions were obtained using
so-called "harmonic function rules" (inspired by supersymmetry
arguments).

In this  paper we consider a general scheme based
on $\sigma$-model approach.
(The $\sigma$-model corresponding to
pure gravitational isotropic model with $n$ Einstein internal spaces
was studied  in  \cite{Ber}-\cite{IM}. Thus, here as in ref.
\cite{IM}  we extend our investigations of multidimensional cosmology
\cite{IM3} to gravitational case.)
In Sect. 2 we describe the model and obtain
its $\sigma$-model representation. In Sect. 3 the non-exceptional
case $d_0 = {\rm dim} M_0 \neq 2$ is considered. In this case the
Einstein-Pauli (EP) frame exists \cite{Ber}-\cite{IM}.
When all internal spaces $M_{i}$ are Ricci-flat and cosmological term
is zero, the  $\sigma$-model  in EP frame has a rather simple
form allowing to obtain certain "Majumdar-Papapetrou (MP) type" solutions
(see Proposition 1) and as a consequence to get a family of "MP-type"
solutions in the considered model (\ref{3.35})-(\ref{3.41}).
The solutions are labelled by
sets  $\Omega_{*} \subset  \Omega$, where $\Omega$ is defined in
(\ref{2.16}). We get restrictions on dilatonic couplings,
dimensions of membranes $d(I) = {\rm dim} M_I$
and its signature parameters $\varepsilon(I) = \pm 1$ (\ref{2.31}),
$I \in \Omega_{*}$.
In Sect. 4 we present the metric in the form
generalizing Tseytlin's  "harmonic function rules". We also
consider some special classes of solutions:
i) in arbitrary dimensions (based on solutions of two Diophantus
equations);
ii)  for $D = 11$ supergravity (e.g. solution with seven Euclidean
intersecting 2-branes).
We also consider the "multi-block" generalization
of the solutions from refs. \cite{V,AVV} containing the
maximal $p$-brane (or monopole in dual version).

\psection{The model}

We consider the model governed by the action
\bearr
\label{2.1}
S = \frac{1}{2\kappa^{2}}
\int_{M} d^{D}z \sqrt{|g|} \{ {R}[g] - 2 \Lambda \nnn
- \sum_{I \in \Omega} [ g^{MN} \partial_{M} \varphi^I
\partial_{N} \varphi^I \nnn
+ \frac{1}{n_I!} \exp(2 \lambda_{JI} \varphi^J) (F^I)^2_g ] \}+ S_{GH},
\ear
where $g = g_{MN} dz^{M} \otimes dz^{N}$ is the metric, $\varphi^I$
is dilatonic scalar field,
\beq
F^I =  dA^I =
\frac{1}{n_I!} F^I_{M_1 \ldots M_{n_I}}
dz^{M_1} \wedge \ldots \wedge dz^{M_{n_I}}     \label{2.2}
\eeq
is $n_I$-form ($n_I \geq 2$) on $D$-dimensional manifold $M$,
$\Lambda$ is a cosmological constant
and $\lambda_{JI} \in {\bf R}$, $I,J \in \Omega$.
In (\ref{2.1})
we denote $|g| = |\det (g_{MN})|$,
\beq    \label{2.3}
(F^I)^2_g =
F^I_{M_1 \ldots M_{n_I}} F^I_{N_1 \ldots N_{n_I}}
g^{M_1 N_1} \ldots g^{M_{n_I} N_{n_I}},
\eeq
$I \in \Omega$,  and $S_{\rm GH}$ is the standard
Gibbons-Hawking boundary term \cite{GH}.
This term is essential for a quantum treatment  of the problem.
Here $\Omega$  is a non-empty finite set.
The action  (\ref{2.1}) with $\Lambda = 0$  and equal $n_I$
was considered recently in \cite{KKLP}. (For supergravity models
with different $n_I$ see also \cite{LPS1})

The equations of motion corresponding to  (\ref{2.1}) have the following
form
\bearr  \label{2.4}
R_{MN} - \frac{1}{2} g_{MN} R  =   T_{MN}
- \Lambda g_{MN}, \\ \lal
\label{2.5} {\btu}[g] \varphi^J -
\sum_{I \in \Omega} \frac{\lambda_{JI}}{n_I!}
\exp(2 \lambda_{KI} \varphi^K) (F^I)^2_g = 0,  \\ \lal
\label{2.6}
\nabla_{M_1}[g](\exp(2 \lambda_{KI} \varphi^K)
F^{I, M_1 \ldots M_{n_I}})  =  0,
\ear
$I,J \in \Omega$.  In (\ref{2.4})
\bearr \label{2.7}
T_{MN} =  \sum_{I \in \Omega} [ T_{MN}^{\varphi^I} +
\exp(2 \lambda_{JI} \varphi^J) T_{MN}^{F^I} ],
\ear
where
\bearr \label{2.8}
T_{MN}^{\varphi^I} =
\p_{M} \varphi^I \p_{N} \varphi^I -
\frac{1}{2} g_{MN} \p_{P} \varphi^I \p^{P} \varphi^I, \\ \lal
T_{MN}^{F^I} = \frac{1}{n_{I}!}  [ - \frac{1}{2} g_{MN} (F^{I})^{2}_{g}
\nnn +
n_{I}
F^{I}_{M M_2 \ldots M_{n_I}} F_{N}^{I, M_2 \ldots M_{n_I}}].
\label{2.9}
\ear
In (\ref{2.5}), (\ref{2.6}) ${\btu}[g]$ and ${\btd}[g]$
are Laplace-Beltrami and covariant derivative operators respectively
corresponding to  $g$.

Let us consider the manifold
\beq  \label{2.10}
M = M_{0}  \times M_{1} \times \ldots \times M_{n},
\eeq
with the metric
\beq      \label{2.11}
g= \e^{2{\gamma}(x)} g^0  +
\sum_{i=1}^{n} \e^{2\phi^i(x)} g^i ,
\eeq
where
\beq    \label{2.12}
g^0  = g^0 _{\mu \nu}(x) dx^{\mu} \otimes dx^{\nu}
\eeq
is a metric on the manifold $M_{0}$ and $g^i  =
g_{m_{i} n_{i}}(y_i) dy_i^{m_{i}} \otimes dy_i^{n_{i}}$ is a metric on
$M_{i}$  satisfying the equation
\beq   \label{2.13}
R_{m_{i}n_{i}}[g^i ] = \lambda_{i} g^i_{m_{i}n_{i}},
\eeq
$m_{i},n_{i}=1,\ldots, d_{i}$; $\lambda_{i}= {\rm const}$,
$i=1,\ldots,n$.  Thus, $(M_i, g^i )$  are Einstein spaces.
The functions $\gamma, \phi^{i} : M_0 \rightarrow {\bf R}$ are smooth.
Here we denote $d_{\nu} = {\rm dim} M_{\nu}$, $\nu = 0, \ldots, n$.

We claim any manifold $M_{\nu}$ to be oriented and connected,
$\nu = 0,\ldots,n$. Then the volume $d_i$-form
\beq    \label{2.14}
\tau_i  = \sqrt{|g^i(y_i)|}
\ dy_i^{1} \wedge \ldots \wedge dy_i^{d_i},
\eeq
and signature parameter
\beq    \label{2.15}
\varepsilon(i)  = {\rm sign}( {\rm det } (g^i_{m_{i}n_{i}})) = \pm 1
\eeq
are correctly defined for all $i=1,\ldots,n$.

Let $\Omega$ from (\ref{2.1}) be a set of all ordered (non-empty)
subsets of
$(1, \ldots,n)$, i.e.
\bearr    \label{2.16}
\Omega = \Omega(n) \equiv \nnn
\{ (1), (2), \ldots, (n), (1,2), \ldots,
(1,2, \ldots, n) \}.
\ear
Clearly, that the number of elements in $\Omega$
is $|\Omega| = 2^n - 1$.

For any $I = (i_1, \ldots, i_k) \in \Omega$, $i_1 < \ldots < i_k$,
we put in (\ref{2.2})
\beq    \label{2.17}
A^I = \Phi^I \tau_{i_1}  \wedge \ldots \wedge \tau_{i_k},
\eeq
where functions $\Phi^I : M_0 \rightarrow {\bf R}$ are smooth,
$I \in \Omega$, and $\tau_{i}$ are defined in (\ref{2.14}).
In components relation (\ref{2.17}) reads
\bearr    \label{2.18}
A^{I}_{P_1 \ldots P_{d(I)}}(x,y) = \nnn
\Phi^{I}(x) \sqrt{|g^{i_1}(y_{i_1})|} \ldots \sqrt{|g^{i_k}(y_{i_k})|}
\ \varepsilon_{P_1 \ldots P_{d(I)}},
\ear
where
\beq    \label{2.19}
d(I) \equiv d_{i_1} + \ldots + d_{i_k}
=  \sum_{i \in I} d_i
\eeq
is dimension of the oriented manifold
\beq  \label{2.20}
M_{I} = M_{i_1}  \times  \ldots \times M_{i_k},
\eeq
and indices $P_1, \ldots, P_{d(I)}$   correspond
to $M_I$.

It follows from (2.17) that
\beq    \label{2.21}
F^I = dA^I = d \Phi^I \wedge \tau_{i_1}  \wedge \ldots \wedge \tau_{i_k},
\eeq
or, in components,
\bearr    \label{2.22}
F^{I}_{\mu P_1 \ldots P_{d(I)}} =
- F^{I}_{P_1 \mu \ldots P_{d(I)}} = \ldots = \nnn
\p_{\mu} \Phi^{I} \sqrt{|g^{i_1}|} \ldots \sqrt{|g^{i_k}|}
\ \varepsilon_{P_1 \ldots P_{d(I)}}.
\ear
It follows from (\ref{2.21}) that
\beq    \label{2.23}
n_I = d(I) + 1,
\eeq
$I \in \Omega$.

\medskip\noi
{\bf Remark 1.} It is more correct to write in
(\ref{2.11}) $\hat{g}^{\alpha}$ instead  $g^{\alpha}$,
where $\hat{g}^{\alpha} =  p_{\alpha}^{*} g^{\alpha}$ is the
pullback of the metric $g^{\alpha}$  to the manifold  $M$
by the canonical projection: $p_{\alpha} : M \rightarrow M_{\alpha}$,
$\alpha = 0, \ldots, n$.
Analogously, we should write $\hat{\Phi}^{I}$ and $\hat{\tau}_{i}$
instead  $\Phi^{I}$ and $\tau_{i}$  in (\ref{2.17}) and
(\ref{2.21}) respectively. In what follows
we omit all "hats" in order to simplify notations.

For dilatonic scalar fields we put
\beq    \label{2.24}
\varphi^I = \varphi^I(x),
\eeq
$I \in \Omega$. Thus, we consider Ansatz when all fields
are essentially depending on points of $M_0$.

The nonzero Ricci tensor components
for the metric (\ref{2.11}) are following \cite{IM}
\bearr   \label{2.25}
R_{\mu \nu}[g]  =   R_{\mu \nu}[g^0 ] +
          g^0 _{\mu \nu} \Bigl[- \Delta_0 \gamma
          +(2-d_0)  (\p \gamma)^2                    \nnn \quad
- \p \gamma \sum_{j=1}^{n} d_j \p \phi^j ]
+ (2 - d_0) (\gamma_{;\mu \nu} - \gamma_{,\mu} \gamma_{,\nu}) \nnn \quad
 - \sum_{i=1}^{n} d_i ( \phi^i_{;\mu \nu} - \phi^i_{,\mu} \gamma_{,\nu}
 - \phi^i_{,\nu} \gamma_{,\mu} + \phi^i_{,\mu} \phi^i_{,\nu}),  \\ \lal
\label{2.26}
R_{m_{i} n_{i}}[g]  = {R_{m_{i} n_{i}}}[g^i ]
     - \e^{2 \phi^{i} - 2 \gamma} g^i _{m_{i} n_{i}}
      \biggl\{ \Delta_0 \phi^{i} \nnn \quad
+ (\p \phi^{i}) [ (d_0 - 2) \p \gamma  +
          \sum_{j=1}^{n} d_j \p \phi^j ] \biggr\},
\ear
Here
$\p \beta \,\p \gamma \equiv g^{0\  \mu \nu} \beta_{, \mu} \gamma_{, \nu}$
and  $\Delta_0$ is the Laplace-Beltrami operator corresponding
to  $g^0 $. The scalar curvature for (\ref{2.11}) is \cite{IM}
\bearr    \label{2.27}
  \nq R[g] =  \sum_{i =1}^{n} \e^{-2 \phi^i} {R}[g^i ]
          + \e^{-2 \gamma} \biggl\{ {R}[g^0 ]
          - \sum_{i =1}^{n} d_i (\p \phi^i)^2  \nnn
 -  (d_0 {-} 2) (\p \gamma)^2
     -  (\p f)^2 - 2 \Delta_0 (f +  \gamma) \biggr\},
\ear
where
\beq \label{2.28}
f = {f}(\gamma, \phi)  = (d_0 - 2) \gamma +\sum_{j=1}^{n} d_j  \phi^j .
\eeq

Using (\ref{2.25}) and (\ref{2.26}),
it is not difficult to verify that the field equations
(\ref{2.4})-(\ref{2.6}) for the field configurations
from  (\ref{2.11}), (\ref{2.21}) and (\ref{2.24})
may be obtained as the
equations of motion corresponding to the action
\bear \label{2.29}
\lal
S_{\sigma} = S_{\sigma}[g^0 , \gamma,\phi,\varphi,\Phi]
\nn \lal =   \frac{1}{2 \kappa^{2}_0}
     \int_{M_0} d^{d_0}x   \sq {g^{0}}
     \e^{f(\gamma, \phi)} \biggl\{ {R}[g^0 ] \nnn
- \sum_{i =1}^{n} d_i (\p \phi^i)^2  -  (d_0 - 2) (\p \gamma)^2
           + (\p f) \p (f + 2\gamma) \nnn
+      \sum_{i=1}^{n} \lambda_{i} d_i \e^{-2 \phi^i + 2 \gamma} -
     2 \Lambda \e^{2 \gamma} - {\cal L}  \biggr\},
\ear
where
\bearr \label{2.30}
{\cal L} = {\cal L}[g^0,\phi,\varphi,\Phi]
= \sum_{I \in \Omega} [ (\p \varphi^I)^2  \nnn
+ \varepsilon(I) \exp(2 \lambda_{JI} \varphi^J -
 2 \sum_{i \in I} d_i \phi^i) (\p \Phi^I)^2 ],
\ear
$|g^{0}|= |\det (g^0 _{\mu\nu})|$ and similar notations are applied
to the metrics $g^{i}$, $i=1, \ldots, n$. In (\ref{2.30})
\beq \label{2.31}
\varepsilon(I) \equiv
\varepsilon(i_1) \times \ldots \times \varepsilon(i_k) = \pm 1
\eeq
for $I = (i_1, \ldots, i_k) \in \Omega$ ($i_1< \ldots < i_k$).

For finite internal space volumes (e.g. compact $M_i$)
\beq   \label{2.32}
V_i = \int_{M_i} d^{d_i}y_i \sq{g^i } < + \infty,
\eeq
the action (\ref{2.29}) (with ${\cal L}$ from (\ref{2.30}) )
coincides with the action (\ref{2.1}), i.e.
\beq   \label{2.33}
  S_{\sigma}[g^0 , \gamma,\phi,\varphi,\Phi] =
  S[g(g^0,\gamma,\phi), \varphi, F(\Phi)],
\eeq
where $g = g(g^0,\gamma,\phi)$  and $F = F(\Phi)$
are defined by the relations (\ref{2.11}) and (\ref{2.21})
respectively and
\beq    \label{2.34}
\kappa^{2} = \kappa^{2}_0 \prod_{i=1}^{n} V_i.
\eeq
This may be readily verified using the following relation for
the scalar curvature (\ref{2.27}) \cite{IM}:
\bearr   \label{2.35}
\nq R[g] {=}  \sum_{i=1}^{n} \! \e^{-2 \phi^i} {R}[g^i ]
          + \e^{-2 \gamma} \Bigl\{ {R}[g^0 ]
          - \sum_{i =1}^{n} d_i (\p \phi^i)^2  \nnn
- (d_0 - 2) (\p \gamma)^2 + (\p f) \p (f + 2 \gamma) + R_{B} \Bigr\},
\ear
where
\bearr \label{2.36}
R_B = (1/\sq {g^0 }) \e^{-f}
     \p_{\mu} [-2 \e^f \sq {g^0 }
     g^{0\ \mu \nu} \p_{\nu} (f + \gamma)]  \nnn
\ear
gives rise to the Gibbons-Hawking boundary term
\beq \label{2.37}
S_{\rm GH} = \frac{1}{2\kappa^{2}} \int_{M} d^{D}z \sq g
     \{ - \e^{-2 \gamma} R_{B} \}.
\eeq
We note that the second term in (\ref{2.30}) appears
due to the following relation (see (\ref{2.22}))
\beq \label{2.38}
\frac{1}{n_I!} (F^I)^2_g
= \varepsilon(I)
\exp(- 2 \gamma  - 2 \sum_{i \in I} d_i \phi^i) (\p \Phi^I)^2.
\eeq

\psection{Exact solutions}

Let us consider the case $d_0 \neq 2$.
In order to simplify the action (\ref{2.29}), we use,
as in ref. \cite{IM},  for $d_0 \neq 2$ the gauge
\beq  \label{3.1}
\gamma = {\gamma}_{0}(\phi) =
\frac{1}{2- d_0}  \sum_{i =1}^{n} d_i \phi^i.
\eeq
It means that $f = {f}(\gamma_0, \phi)= 0$, or the conformal
Einstein-Pauli frame is used. (This frame does not exist for $d_0 = 2$.)
For the cosmological case  $d_0 =1$,  $g^0  = - dt \otimes dt$
and (\ref{3.1}) corresponds to the harmonic-time gauge  \cite{IM1,IMZ}.
>From (\ref{2.29}), (\ref{2.30}), (\ref{3.1}) we get
\bearr  \label{3.2}
S_{0}[g^0 ,\phi, \varphi,\Phi] =
S_{\sigma}[g^0,\gamma_0(\phi),\phi,\varphi,\Phi] =  \nnn
\frac{1}{2 \kappa^{2}_0}
     \int_{M_0} d^{d_0}x \sq {g^0 } \Bigl\{ {R}[g^0 ] \nnn
- G_{ij} g^{0\ \mu \nu} \p_{\mu} \phi^i  \p_{\nu} \phi^j -
2 {V}(\phi) - \sum_{I \in \Omega} \Bigl[ (\p \varphi^I)^2  \nnn
+ \varepsilon(I) \exp(2 \lambda_{JI} \varphi^J -
 2 \sum_{i \in I} d_i \phi^i) (\p \Phi^I)^2 \Bigr] \Bigr\},
\ear
where
\beq     \label{3.3}
G_{ij} = d_i \delta_{ij} + \frac{d_i d_j}{d_0 -2}
\eeq
are the components of the ("pure gravitational")
"midisuperspace"  metric on
${\bf R}^{n}$  \cite{IM}
(or gravitational part of target space metric), $i, j = 1, \ldots, n$,
and
\beq  \label{3.4}
\nq V = {V}(\phi) = \Lambda \e^{2 {\gamma_0}(\phi)} - \half
     \sum_{i =1}^{n} \lambda_i d_i \e^{-2 \phi^i + 2 {\gamma_0}(\phi)}
\eeq
is the potential. Thus, we are led to the action of a self-gravitating
$\sigma$ model on  $M_0$
with a  $(n + 2 |\Omega|)$-dimensional target space  ($|\Omega| = 2^n - 1$)
and a self-interaction described by the
potential (\ref{3.4}).

\subsection{$\sigma$-model with zero potential.}

Now we consider the case $\lambda_i = \Lambda = 0$,
i.e. all spaces $(M_i, g^i)$ are Ricci-flat,
$i = 1, \ldots, n$, and cosmological constant is zero.
In this case the potential  (\ref{3.4}) is trivial $V =0$
and we are led to the $\sigma$-model with the following
action.
\bearr \label{3.5}
S_{\sigma} =  \int_{M_0} d^{d_0}x \sq {g^0 } \Bigl\{ {R}[g^0]
- \hat{G}_{AB} \p \sigma^A \p \sigma^B  \nnn
- \sum_{I \in \Omega} \varepsilon(I) \exp(2 L_{AI} \sigma^A)
(\p \Phi^I)^2  \Bigr\},
\ear
where we put $2 \kappa^{2}_0 =1$. In (\ref{3.5})
$(\sigma^A) = (\phi^j, \varphi^J) \in {\bf R}^{N}$,
where $N = n + m$, $m = |\Omega| = 2^n - 1$,
\beq \label{3.6}
\hat{G} = \left(\hat{G}_{AB} \right)    =
                             \left(
                              \begin{array}{cc}
                                G_{ij} &  0 \\
                                   0   &  \delta_{IJ}
                               \end{array}
                          \right)
\eeq
is non-degenerate (block-diagonal) $N \times N$-matrix and
\beq \label{3.7}
L = \left(L_{AI} \right)    =
                             \left(
                              \begin{array}{cc}
                               &l_{jI} \\
                               &\lambda_{JI}
                               \end{array}
                          \right)
\eeq
is $N \times m$-matrix with
\beq \label{3.8}
l_{jI} = - \sum_{i \in I} d_i \delta^i_j,
\eeq
$i,j = 1, \ldots, n$, $I,J \in \Omega$.

The equations of motion corresponding to the $\sigma$-model
action (\ref{3.5}) are  following
\bearr  \label{3.9}
R_{\mu \nu}[g^0]  =
\hat{G}_{AB} \p_{\mu} \sigma^A \p_{\nu} \sigma^B  \nnn
+ \sum_{I \in \Omega} \varepsilon(I) \exp(2 L_{AI} \sigma^A)
\p_{\mu} \Phi^I  \p_{\nu} \Phi^I,   \\ \lal
\label{3.10}
\hat{G}_{AB} {\btu}[g^0] \sigma^B \nnn -
 \sum_{I \in \Omega} \varepsilon(I) L_{AI}
\exp(2 L_{AI} \sigma^A) (\p \Phi^I)^2 = 0,  \\ \lal
\label{3.11}
\p_{\mu} \left( \sqrt{|g^0|}
g^{0 \mu \nu} \exp(2 L_{AI} \sigma^A) \p_{\nu} \Phi^I \right) = 0,
\ear
$A = 1, \ldots, N$; $I \in \Omega$.

Now we present a special class of solutions to field equations
(\ref{3.9})-(\ref{3.11}) (of Majumdar-Papapetrou type).

{\bf Proposition 1.} Let  $\Omega_{*} \subset \Omega$ be a non-empty
set of indices such that there exist a set of real non-zero
numbers $\nu_I, I \in \Omega_{*}$  ($\nu_I \in {\bf R} \setminus \{0 \}$)
satisfying the relations
\beq \label{3.12}
 (L^{T} \hat{G}^{-1} L)_{IJ} = - \varepsilon(I) (\nu_I)^{-2}\delta_{IJ},
\eeq
$I,J \in \Omega_{*}$.
Then the following field configuration
\bearr  \label{3.13}
R_{\mu \nu}[g^0]  = 0, \\ \lal
\label{3.14}
\sigma^A = \sum_{I \in \Omega_{*}} \alpha^A_I \ln H_I + \sigma^A_0, \\ \lal
\label{3.15}
\Phi^I  = \frac{\nu_I \exp(- L_{AI} \sigma^A_0)}{H_I}, \quad
I \in \Omega_{*}, \\ \lal
\label{3.16}
\Phi^{\bar{I}} = {\bar{C}}_{\bar{I}} = {\rm const},  \qquad
\bar{I} \in  \Omega \setminus  \Omega_{*}
\ear
$\sigma^A_0 =$ const, $A = 1, \ldots, N$, satisfies
the field equations (\ref{3.9})-(\ref{3.11}) if
\beq \label{3.17}
\alpha^A_I = -  \hat{G}^{AB} L_{BI} \varepsilon(I) (\nu_I)^{2},
\eeq
($A = 1, \ldots, N$; $I \in \Omega_{*}$) $\nu_I$
satisfy (\ref{3.12}) and functions $H_I = H_{I}(x) > 0$ are
harmonic, i.e.
\beq  \label{3.18}
{\btu}[g^0] H_I = 0,
\eeq
$I \in \Omega_{*}$.

In (\ref{3.17}) we use the notation $(\hat{G}^{AB}) =
(\hat{G}_{AB})^{-1}$.  The Proposition 1 may by proved
immediately by a substitution of (\ref{3.12})-(\ref{3.18}) into
the  equations of motion (\ref{3.9})-(\ref{3.11}).

Note that the relation (\ref{3.12}) may be written in the
following form
\beq \label{3.19}
 (L_{I}, L_{J}) = - \varepsilon(I) (\nu_I)^{-2}\delta_{IJ},
\eeq
$I,J \in \Omega_{*}$, where
\beq  \label{3.20}
L_{I} = (L_{AI}) \in {\bf R}^N
\eeq
and
\beq
\label{3.21}
(X,Y) \equiv X_A \hat{G}^{AB} X_{B}
\eeq
is non-degenerate (real-valued) quadratic form. Thus,
due to (\ref{3.19}) the set of
vectors $(L_I$, $I \in \Omega_{*}$) is orthogonal one
and any $(L_{I}, L_{I})$ have the opposite sign to
$\varepsilon(I)$, $I \in \Omega_{*}$. Clearly, if quadratic form
(\ref{3.21}) is positively definite (for our model this takes place
for $d_0 > 2$), then $\varepsilon(I) < 0$ for all $I \in \Omega_{*}$.

Now, we apply the Proposition 1 to the considered here
model with Ricci-flat spaces  $(M_i, g^i)$, $i = 1, \ldots, n$,
and zero cosmological constant.  From (\ref{3.6}), (\ref{3.7}) and
(\ref{3.21}) we get
\beq \label{3.22}
(L_{I}, L_{J}) =  \avers{l_I, l_J }  + \vec{\lambda}_I \vec{\lambda}_J,
\eeq
$I,J \in \Omega$, where vectors
\beq \label{3.23}
l_I = (l_{jI}) \in {\bf R}^n
\eeq
are defined in  (\ref{3.8})  and
\beq \label{3.24}
\vec{\lambda}_I = (\lambda_{JI}) \in {\bf R}^m,
\eeq
$I \in \Omega$ ($m = |\Omega|$).
In (\ref{3.22})
\beq \label{3.25}
\avers{u,v} \ \equiv    u_i G^{ij} v_j
\eeq
is a quadratic form on  ${\bf R}^n$. Here, as in \cite{IM},
\beq \label{3.26}
G^{ij} =
\frac{\delta_{ij}}{d_i} + \frac{1}{2 - D}
\eeq
are components of the matrix inverse to the matrix  $(G_{ij})$ in
(\ref{3.3}).

The calculation gives us the relation
\beq \label{3.27}
\avers{l_I, l_J} = d(I \cap J) + \frac{d(I) d(J)}{2-D},
\eeq
$I, J \in \Omega$.
Here, like in relation  $i \in I$, we identify
the ordered set $I = (i_1, \ldots, i_k)$, $i_1 < \ldots < i_k$,
with the corresponding set  $\{ i_1, \ldots, i_k \}$.
In (\ref{3.27}) $d(\emptyset) = 0$.

Due to (\ref{3.19}), (\ref{3.22}) and (\ref{3.27})  the
relation (\ref{3.12}) reads
\beq \label{3.28}
d(I \cap J) + \frac{d(I) d(J)}{2-D} + \vec{\lambda}_I \vec{\lambda}_J
= - \varepsilon(I) (\nu_I)^{-2}\delta_{IJ}
\eeq
$I, J \in \Omega_{*}$.

For coefficients $\alpha^A_I$ from (\ref{3.17}) we get
\bearr  \label{3.29}
\alpha^i_I  = \biggl( \sum_{j \in I} \delta^i_j  +
\frac{d(I)}{2-D} \biggr) \varepsilon(I) (\nu_I)^{2},  \\ \lal
\label{3.30}
\alpha^J_I = - \lambda_{JI}  \varepsilon(I) (\nu_I)^{2},
\ear
$I \in \Omega_{*}$, $J \in \Omega$, $i = 1, \ldots, n$.

Relations (\ref{3.14}) ($(\sigma^A) = (\phi^j, \varphi^J)$)
read
\bearr  \label{3.31}
\phi^i = \sum_{I \in \Omega_{*}} \alpha^i_I   \ln H_I  +  \phi^i_0,
  \\ \lal \label{3.32}
\varphi^J = \sum_{I \in \Omega_{*}} \alpha^J_I   \ln H_I
+  \varphi^J_0,
\ear
$J \in \Omega$; $i = 1, \ldots, n$.  These relations
imply for $\gamma$ from  (\ref{3.1}) \beq  \label{3.33} \gamma = \sum_{I
\in \Omega_{*}} \alpha^0_I  \ln H_I  +  \gamma_0(\phi_0), \eeq where \beq
\label{3.34} \alpha^0_I  = \frac{d(I)}{2-D} \varepsilon(I) (\nu_I)^{2},
\eeq
$I \in \Omega_{*}$.

\subsection{The solution.}

Thus we obtained the following solutions to
equations of motion (\ref{2.4})-(\ref{2.6}) with  $\Lambda = 0$
defined on the manifold (\ref{2.10}):
\bearr         \label{3.35}
g= c_0^2 \biggl( \prod_{I \in \Omega_{*}} H_I^{2 \alpha^0_I} \biggr)
g^0 + \sum_{i=1}^{n} c_i^2
\biggl( \prod_{I \in \Omega_{*}} H_I^{2 \alpha^i_I} \biggr)  g^i
\\ \lal
\label{3.36}
{\rm Ric}[g^0]= {\rm Ric}[g^1] = \ldots = {\rm Ric}[g^n] = 0, \\ \lal
\label{3.37}
\varphi^J = - \sum_{I \in \Omega_{*}}
\lambda_{JI}  \varepsilon(I) (\nu_I)^{2}  \ln H_I  +  \varphi^J_0,
\\
\lal
\label{3.38}
A^I = \frac{\nu_I }{D_I H_I} \tau_I, \ I \in \Omega_{*},  \\ \lal
\label{3.39}
A^{\bar{I}} = \bar{C}_{\bar{I}} \tau_{\bar{I}}, \qquad
\bar{I} \in  \Omega \setminus  \Omega_{*}
\ear
where  $c_0, c_i, D_I > 0$,
$\varphi^J_0$, $\bar{C}_{\bar{I}}$ are
constants satisfying
\bearr \label{3.40}
c_0^{2-d_0} = \prod_{i=1}^{n} c_i^{d_i}, \\ \lal
\label{3.41}
D_I = \prod_{i \in I} c_i^{-d_i} \exp(\lambda_{JI} \varphi^J_0),
\ear
$\tau_I = \tau_{i_1}  \wedge \ldots \wedge \tau_{i_k}$,
for $I = (i_1, \ldots, i_k)$, $\tau_{i}$ are defined in (\ref{2.14}),
parameters $\nu_I \neq 0$, $\vec{\lambda}_I = (\lambda_{JI})$
and $\varepsilon(I) = \pm 1$ (see (\ref{2.31}) ) satisfy the relation
(\ref{3.28}), parameters $ \alpha^i_I$, $\alpha^0_I$ are defined
in (\ref{3.29}), (\ref{3.34}),  and  functions
$H_I = H_{I}(x) > 0$,
are harmonic on $(M_0, g^0)$, i.e.
${\btu}[g^0] H_I = 0$, $I \in \Omega_{*}$.  In (\ref{3.36})
${\rm Ric}[g^{\nu}]$ is Ricci-tensor corresponding to $g^{\nu}$.

{\bf $d_0 =2$ case}.
We note that, although the presented here solution was obtained
for $d_0 \neq 2$, it is valid also for $d_0 = 2$. It may be verified
by a straightforward substitution of presented above relations
into equations of motions  or  using the $\sigma$-model
representation (\ref{2.29}) for $d_0 = 2$.

Due to (\ref{3.37})
\beq
\label{3.42}
\exp ( 2 \vec{\lambda}_{I'}  (\vec{\varphi}
- \vec{\varphi}_0) ) =   \prod_{I \in \Omega_{*}}
H_I^{- 2 \varepsilon(I) \nu_I^{2}
\vec{\lambda}_{I'} \vec{\lambda}_{I}},
\eeq
$I' \in \Omega$, where $\vec{\varphi} = (\varphi^I)$.

{\bf Proposition 2}. For $d_0 \geq 2$  eqs. (\ref{3.28})  imply
\beq
\label{3.43}
       \varepsilon(I) = -1
\eeq
for all $I \in \Omega_{*}$.

{\bf Proof.} Let us suppose that $\varepsilon(I) = 1$ for some
$I \in \Omega_{*}$. Then due to (\ref{3.28})
\bearr
\label{3.44}
d(I) + \frac{(d(I))^2}{2-D} \nnn
\quad = \frac{d(I)}{D-2} (D -2 -d(I)) < 0
\ear
or, equivalently,
\beq
\label{3.45}
D - 2 - d(I) = d_0 - 2 + d(I_0) - d(I) < 0,
\eeq
where here and below
\beq
\label{3.46}
I_0 \equiv \{1, \ldots, n \},
\eeq
is the maximal element in $\Omega$. But due to $d(I_0) - d(I) \geq 0$ and
$d_0 \geq 2$, (\ref{3.45}) is never satisfied. This proves the
proposition.

{\bf Remark 2.} It follows from (\ref{3.28}) and (\ref{3.45}) that
$\varepsilon(I) = 1$ may take place only if: $d_0 = 1$, $I = I_0$
and $\vec{\lambda}_I^2 < (D-1)/(D-2)$.

Thus, it follows from Proposition 1 and Remark 2 that the
case of negative $\varepsilon(I)$ is rather general one.

\psection{Special solutions}

In this section we consider the solutions with  $\varepsilon(I) = -1$
for all $I \in \Omega_{*}$. From (\ref{3.29}), (\ref{3.34})
 and (\ref{3.35}) we get in
this case
\bearr      \label{4.1}
g=  \biggl( \prod_{I \in \Omega_{*}} H_I^{- 2 d(I) \nu^2_I}
\biggr)^{1/(2-D)} \times \nnn
\times \biggl\{ g^0 +
\sum_{i=1}^{n} \biggr( \prod_{I \in
\Omega_{*}, \ I \ni i} H_I^{- 2 \nu^2_I} \biggr)  g^i \biggr\},
\ear

Here we put for simplicity $c_0 = c_i = 1$  in (\ref{3.35}).
(As usual $\prod_{\emptyset} \ldots \equiv 1$). Relations
(\ref{3.28}) read
\bearr
\label{4.2}
(\nu_I)^{-2} =
d(I) + \frac{(d(I))^2}{2-D} + \vec{\lambda}_I^2, \\ \lal
\label{4.3}
d(I \cap J) + \frac{d(I) d(J)}{2-D} +
\vec{\lambda}_I \vec{\lambda}_J =0, \  \  I \neq J,
\ear
$I,J \in \Omega_{*}$.

\subsection{The case of orthogonal $\vec{\lambda}_I$}.

Here we consider the case, when
\beq      \label{4.4}
   \vec{\lambda}_I \vec{\lambda}_J =0, \qquad  I \neq J,
\eeq
$I,J \in \Omega_{*}$.
>From (\ref{4.3}) we get
\beq
\label{4.5}
d(I \cap J) = \frac{d(I) d(J)}{D-2}, \qquad I \neq J,
\eeq
$I,J \in \Omega_{*}$.

Here an interesting problem arises: for given $n$ and
$d_0, d_1, \ldots, d_n \in {\bf N}$  to find $\Omega_{*} \subset \Omega$
such that (\ref{4.5}) is satisfied.

{\bf Remark 3.} It follows from (\ref{4.5}) that $I \cap J \neq
\emptyset$ for all  $I,J \in \Omega_{*}$, i.e. any two membranes are
intersecting.  In $D = \infty$  case, when all sets are finite, we obtain
$I \cap J = \emptyset$, $I \neq J$, $I,J \in \Omega_{*}$
(membranes are non-intersecting).

Here we consider a simple configuration for $n \geq 3$:
\bearr     \label{4.6}
   d_1 = \ldots = d_{n-1} = d, \\ \lal
   \label{4.7}
   \Omega_{*} = \{ (1,n), \ldots , (n-1, n) \},  \\ \lal
   \label{4.8}
   \varepsilon(1) = \ldots = \varepsilon(n-1) = - \varepsilon(n).
   \ear
(The condition (\ref{3.43}) is satisfied in this case.
Then (\ref{4.5}) reads
\beq
\label{4.9}
d_n = \frac{(d+ d_n)^2}{D-2}.
\eeq
This equation with the relation for total dimension
\beq
\label{4.10}
D = d_0 + (n-1) d + d_n
\eeq
form a set of Diophantus equations.

{\bf Proposition 3}. The set of Diophantus equations (\ref{4.9}) and
(\ref{4.10}) has a solution if and only if in prime number decomposition of
$D-2$:
\beq
\label{4.11}
D - 2 = p_1^{k_1}  \ldots p_l^{k_l},
\eeq
there exists a power $k_i \geq 2$  (for some $i \in \{1, \ldots, l \}$).

{\bf Proof. \ Necessity.} Let us suppose an opposite: all $k_i =1$ .
Then $(d + d_n)^2$ and hence $d + d_n$  is divided by any
$p_i$, $i \in \{1, \ldots, l \}$ or,
equivalently,
\beq \label{4.12}
d + d_n = r p_1  \ldots p_l = r (D-2)  \rightarrow d_n = r^2 (D-2),
\eeq
$r \in {\bf N}$. Clearly, that (\ref{4.12}) is impossible.

{\bf Sufficiency.} Here we present the solution of eqs. (\ref{4.9}),
(\ref{4.10}).  Let
\beq
\label{4.13}
\omega_2 = \omega_2 (D -2) \equiv
\{ i \in \{1, \ldots, l \}  | k_i \geq 2 \}
\eeq
($\omega_2 \neq \emptyset$).
We denote
\beq
\label{4.14}
Ò^s \equiv \prod_{\alpha \in \omega} p_{\alpha}^{s_{\alpha}},
\eeq
where  $\omega  \subset \omega_2$ and powers $s_{\alpha}$ satisfy to
relations
\beq
\label{4.15}
1 \leq s_{\alpha} \leq \left[ \frac{k_{\alpha}}{2} \right],
\eeq
$\alpha \in \omega$ ($[x]$ is integer part of $x$). Then, the dimensions
\bearr
\label{4.16}
d_n = (D-2)/ (p^s)^2,   \\ \lal
\label{4.17}
d = d_n (p^s -1),   \\ \lal
\label{4.18}
d_0 - 2 = d (p^s - n + 2)
\ear
($3 \leq n \leq p^s + 2$ for $d_0 \geq 2$)
satisfy  eqs. (\ref{4.9}), (\ref{4.10}). The proposition is
proved.

The dimensions allowed by Proposition 3 are
\beq
\label{4.19}
D = 6, 10, 11, 14, 18, 20, 22, 26, 27, \ldots.
\eeq

The solution (\ref{4.1}) for dimensions from (\ref{4.16})-(\ref{4.18})
has the following form
\beq
\label{4.20}
g= H^{1/p^s} \{ g^0 +  \sum_{i=1}^{n - 1} H_i^{-
2 \nu^2_i}  g^i  + H^{-1}   g^n \},
\eeq
where $H_i = H_{(i,n)}$, $\nu_i = \nu_{(i,n)}$,
\beq
\label{4.21}
H = \prod_{i=1}^{n-1} H_i^{ 2 \*nu^2_i}, \qquad
\nu^2_i  = (d + \vec{\lambda}_i^2 )^{-1}
\eeq
and
$\vec{\lambda}_i = \vec{\lambda}_{(i,n)}$,
$i = 1, \ldots, n -1$.

Now we consider few examples.

{\bf D = 6.} From (\ref{4.16})-(\ref{4.18}) we get
\beq
\label{4.22}
p^s = 2, \  \ d = d_n = 1,   \
d_0 = 6 - n,
\eeq
$n = 3,4$.

In this case all $d(I) = d + d_n =2$,  $I \in \Omega_{*}$, and we
have $(n-1)$ 3-forms generating $(n-1)$ intersecting membranes. The case
$d_0 =3, n = 3$,
$g^0 = dx^{1} \otimes dx^{1} +
dx^{2} \otimes dx^{2} + dx^{3} \otimes dx^{3}$,
$g^3 = - dt \otimes dt$
is of special interest for "physical applications" ($x^i$ are space
coordinates, and $t$ is time).

{\bf D = 10.} We get
\beq
\label{4.23}
p^s = 2, \ \ d = d_n = 2,   \
d_0 = 10 - 2 n,
\eeq
$n = 3,4, 5$.

{\bf D = 11.} In this case
\beq
\label{4.24}
p^s = 3, \ d = 2, \ d_n = 1,   \
d_0 = 12 - 2 n,
\eeq
$n = 3, 4, 5$. For $\vec{\lambda}_i = 0$ we obtain from (\ref{4.20})
and (\ref{4.21})
\bearr
\label{4.25}
g= H^{1/3} \{ g^0 + \sum_{i=1}^{n -1} H_i^{- 1}  g^i  + H^{-1}   g^n \},
\ear
where  $H = \prod_{i=1}^{n-1} H_i$.
This solution with $g^n = - dt \otimes dt$ and flat Euclidean metrics
$g^0, \ldots , g^{n-1}$  was considered in \cite{Ts1} ($n= 3,4$),
\cite{GKT} ($n= 2,3,4,5$) and  \cite{LPS2} ($n= 5$).

\subsection{Systems with maximal element}

Let us consider another example
\beq
\label{4.26}
\Omega_{*} = \{ (1), \ldots , (n_1), I_0 \},
\eeq
where $I_0 =  \{1, \ldots, n \}$ is maximal element and
\ba
\label{4.27}
     n_1 = &n -1,  \quad &n = 2k, \nonumber \\
        = &n,  \qquad &n = 2k -1,
\ea
$k \in {\bf N}$.  We put
\beq
 \label{4.28}
 \varepsilon(1) = \ldots = \varepsilon(n_1) = - 1
 \eeq
and $\varepsilon(n) =  1$ if $n =2k$. In this case all $\varepsilon(I) =
- 1$, $I \in \Omega_{*}$. For parameters   $\nu_i = \nu_{(i)}$,
$\nu_0 = \nu_{I_0}$,
$\vec{\lambda}_i = \vec{\lambda}_{(i)}$,
$\vec{\lambda}_0 = \vec{\lambda}_{I_0}$,
we have from (\ref{4.2})
\bearr
 \label{4.29}
(\nu_i)^{-2} =
d_i + \frac{d_i^2}{2-D} + \vec{\lambda}_i^2, \\ \lal
 \label{4.30}
(\nu_0)^{-2} =
\frac{(D-d_0)(d_0-2)}{D-2} + \vec{\lambda}_0^2,
\ear
$i = 1, \ldots, n_1$. The constraints on $\vec{\lambda}_0$,
$\vec{\lambda}_i$  read (see (4.3))
\bearr
\label{4.31}
 \frac{d_i d_j}{2-D} + \vec{\lambda}_i \vec{\lambda}_j =0,
 \quad  i \neq j,  \\ \lal
\label{4.32}
 \frac{d_i (d_0-2)}{D-2} + \vec{\lambda}_0 \vec{\lambda}_i =0,
\ear
$i,j = 1, \ldots, n_1$.

>From (\ref{4.1}) we get the solution for the metric
\bearr
\label{4.33}
g= H^{1/(2-D)} \{ g^0 +  H_0^{- 2 \nu^2_0} \times \nnn
\times \quad [ \sum_{i=1}^{n_1} H_i^{- 2 \nu^2_i}  g^i
+ (1 - \delta_{n}^{n_1}) g^n ] \},
\ear
where
\beq
\label{4.34}
H = H_0^{- 2 (D - d_0) \nu^2_0}  \prod_{i=1}^{n_1} H_i^{- 2 d_i \nu^2_i}.
\eeq
Here $H_i = H_{(i)}$, $H_0 = H_{I_0}$,
$i = 1, \ldots, n_1$.

The solution (\ref{4.33}) with $n=2$ (and one scalar field)
was considered recently in \cite{AVV} (see also \cite{V} for
special parameters).
Some special cases  of the $n=2$ solution were considered
also in  \cite{KLO} ($D=4$), \cite{Ts2,CM,CH}
($D=10, d_0 =2$).

We also note, that in ref. \cite{IM2} the solution
(\ref{4.33}) with $n= 2$, $d_1 = 1$, $H_0 = 1$,
$g^1 = - dt \otimes dt$ and flat Euclidean metric $g^0$ was
generalized to the case of non-zero cosmological constant $\Lambda$.
The solution \cite{IM2} for certain couplings describes a collection of
multidimensional extremal dilatonic black holes "living"
in expanding Universe  (in the presence
of Ricci-flat internal space).

\subsection{Dual representation. Monopole}

Any  $F^I$-term in the action (\ref{2.1})
may be presented in the following form
\beq
\label{4.35}
\exp(2 \vec{\lambda}_{I} \vec{\varphi} ) (F^I)^2 =
\exp(- 2 \vec{\lambda}_{I} \vec{\varphi} ) (\star F^I)^2 \varepsilon_g,
\eeq
where
\beq
\label{4.36}
\star F^I \equiv
\exp(2 \vec{\lambda}_{I} \vec{\varphi} ) (* F^I),
\eeq
$*$  is Hodge operator corresponding to the metric $g$ and
$\varepsilon_g = {\rm sign }({\rm det}(g_{MN})) = \pm 1$.
The equations of motion corresponding to $F^I$ read
as Bianchi  identities for $\star F^I$ and vice versa
\bearr
\label{4.37}
d (\exp(2 \vec{\lambda}_{I} \vec{\varphi} ) * F^I) = 0
\Leftrightarrow d (\star F^I) = 0, \\ \lal
\label{4.38}
d (\exp(- 2 \vec{\lambda}_{I} \vec{\varphi} ) * (\star F^I)) = 0
\Leftrightarrow d  F^I = 0,
\ear
(the representation $\star F^I = d \bar{A}^I$ exists at least locally)
$I \in \Omega_{*}$.

The term $\star F^{I_0}$  ($I_0 = \{ 1, \ldots, n \}$) is the form of rank
$d_0 -1$ and usually is interpreted as monopole. The calculation gives
\bearr
\label{4.39}
(\star F^{I_0})_{\mu_1 \ldots \mu_{d_0 - 1}}
= \exp ( 2 \vec{\lambda}_{I_0} \vec{\varphi} + 2(d_0 -2) \gamma )
\varepsilon(I_0) \times
\nnn
\times \sqrt{|g^{0}|}
 \varepsilon_{\mu_1 \ldots \mu_{d_0}}.
\btd^{\mu_{d_0}}[g^0] \Phi^{I_0}
\ear
(see relations (\ref{3.33}) and (\ref{3.42}) ).

\subsection{$D = 11$ supergravity}

Let us consider the case: $D = 11$,  $\vec{\lambda}_I =0$,
$I \in \Omega$. From  (\ref{4.2}), (\ref{4.3}) we have the
following possibilities
\beq
\label{4.40}
\{ d(I), d(J) \}  = \{ 3, 3 \},  \{ 3, 6 \},  \{ 6, 6 \}
\eeq
if $d(I \cap J) = 1, 2, 4$ respectively and $2 \nu_I^2 = 1$,
$I,J \in \Omega_{*}$. Using the rules  (\ref{4.40})   we may list
all possible sets  $\Omega_{*}$, or collections of overlapping
2-branes and 5-branes  \cite{Ts1,PT,GKT}.
In this case the relation for the metric (\ref{4.1}) is nothing
more than Tseytlin's "harmonic function rule" \cite{Ts1} obtained
from supersymmetry arguments.

The  4-form
\beq
\label{4.41}
{\cal F}_4 =
\sum_{I \in \Omega_*, \ d(I) = 3}   F^I +
\sum_{I \in \Omega_*, \ d(I) = 6} * F^I,
\eeq
satisfies the field equations of $D = 11$ supergravity
\beq
\label{4.42}
d * {\cal F}_4 =   {\cal F}_4 \wedge  {\cal F}_4
\eeq
due to relation  ${\cal F}_4 \wedge  {\cal F}_4 = 0$,
equations of motion and Bianchi identities for $F^I$,
$I \in \Omega_{*}$.

We note, that all solutions in $D = 11$ supergravity
with overlapping 2- and 5-branes ($d(I) = 3, 6$), considered
in literature (see \cite{Ts1,PT,GKT} and refs. therein),
may be readily incorporated into considered scheme.
For example, three 5-brane solution \cite{Ts1} corresponds to $n =4$,
 $d_0 = 3$, $d_1 = d_2 = d_{3} = d_{4} = 2$,
$\varepsilon(1) = \varepsilon(2) =
\varepsilon(3) = - \varepsilon(4) = 1$,
$\Omega_{*} = \{ (2,3,4), (1,3,4), (1,2,4) \}$.

Here we present an interesting special solution of
$D =11$ supergravity
\bearr
\label{4.43}
g=  \biggl( \prod_{I \in \Omega_{*}} H_I \biggr)^{1/3} \times
\nnn \times
\bigg\{ g^0
- \sum_{i=1}^{7} \biggl( \prod_{I \in
\Omega_{*}, \ I \ni i} H_I^{- 1} \biggr)  dy^{i} \otimes dy^{i}  \biggr\},
\ear
defined on the manifold $M_0 \times {\bf R}^7$. In
(\ref{4.43})  $(M_0, g^0)$ is Ricci-flat 4-dimensional
manifold of signature $(+,-, -, -)$  and
\bearr
\label{4.44}
\Omega_{*} = \{ (1,3,5), (1,2,6), (2,3,4), \nnn
\qquad          (1,4,7), (2,5,7), (3,6,7), (4,5,6) \}
\ear
(Clearly that the condition  (\ref{3.43}) is satisfied in this
case.) This solution describes seven intersecting Euclidean membranes.

\Acknow
{The authors are grateful to K.A. Bronnikov, for useful discussions.}

This work was supported in part by the Russian State Committee for
Science and Technology and Russian Fund for Basic Research.

\end{document}